\documentclass{article}

\usepackage{times}
\usepackage{graphicx}
\usepackage[superscript]{cite}

\newcommand{\ms}{ms$^{-1}$}

\newcommand{\apjs}{Astrophys.~J. Suppl.}
\newcommand{\aap}{Astron. Astrophys.}
\newcommand{\aj}{Astron~J.}
\newcommand{\apj}{Astrophys. J.}
\newcommand{\mnras}{Mon. Not. R. Astron. Soc}
\newcommand{\pasp}{Proc. Astron. Soc. Pacific}
\newcommand{\nat}{Nature}
\newcommand{\actaa}{AcA}

\newcounter{firstbib}

\title{A terrestrial planet candidate in a temperate orbit around Proxima Centauri}

\author{
Guillem            Anglada-Escud\'e$^{1\ast}$,
Pedro~J.           Amado$^{2}$,
John               Barnes$^{3}$,                    \\
Zaira              M. Berdi\~nas$^{2}$,              
R.~Paul            Butler$^{4}$,                        
Gavin~A.~L.        Coleman$^{1}$,                   \\
Ignacio            de~la~Cueva$^{5}$,               
Stefan             Dreizler$^{6}$,
Michael            Endl$^{7}$,                      \\
Benjamin           Giesers$^{6}$,                   
Sandra~V.          Jeffers$^{6}$,                    
James~S.           Jenkins$^{8}$,                   \\
Hugh~R.~A.         Jones$^{9}$,                      
Marcin             Kiraga$^{10}$,                    
Martin             K\"urster$^{11}$,                \\
Mar\'ia~J.         L\'opez-Gonz\'alez$^{2}$,            
Christopher~J.     Marvin$^{6}$,                    
Nicol\'as          Morales$^{2}$,                   \\
Julien             Morin$^{12}$,                    
Richard~P.         Nelson$^{1}$,                       
Jos\'e~L.          Ortiz$^{2}$,                     \\
Aviv               Ofir$^{13}$,                     
Sijme-Jan          Paardekooper$^{1}$,
Ansgar             Reiners$^{6}$,                   \\       
Eloy               Rodr\'iguez$^{2}$,            
Cristina           Rodr\'iguez-L\'opez$^{2}$,           
Luis~F.            Sarmiento$^{6}$,                 \\                 
John~P.           Strachan$^1$,       
Yiannis            Tsapras$^{14}$,             
Mikko              Tuomi$^{9}$,                     \\    
Mathias            Zechmeister$^{6}$.
\\
}
\begin{document}
\baselineskip15pt
\maketitle
\noindent
$^{1}$School of Physics and Astronomy, Queen Mary University of London, 327 Mile End Road, London E1 4NS, UK
\newline
$^{2}$Instituto de Astrofísica de Andalucía - CSIC, Glorieta de la Astronomía S/N, E-18008 Granada, Spain
\newline
$^{3}$Department of Physical Sciences, Open University, Walton Hall, Milton Keynes MK7 6AA, UK
\newline 
$^{4}$Carnegie Institution of Washington, Department of Terrestrial Magnetism
5241 Broad Branch Rd. NW, Washington, DC 20015, USA
\newline
$^{5}$Astroimagen, Ibiza, Spain
\newline
$^{6}$Institut f\"ur Astrophysik, Georg-August-Universit\"at G\"ottingen
Friedrich-Hund-Platz 1, 37077 G\"ottingen, Germany
\newline
$^{7}$The University of Texas at Austin and Department of Astronomy and McDonald Observatory
2515 Speedway, C1400, Austin, TX 78712, USA
\newline
$^{8}$Departamento de Astronomía, Universidad de Chile 
Camino El Observatorio 1515, Las Condes, Santiago, Chile
\newline
$^{9}$Centre for Astrophysics Research, Science \& Technology Research Institute, 
University of Hertfordshire, Hatfield AL10 9AB, UK
\newline
$^{10}$Warsaw University Observatory, Aleje Ujazdowskie 4, Warszawa, Poland
\newline
$^{11}$Max-Planck-Institut f\"ur Astronomie 
K\"onigstuhl 17, 69117 Heidelberg, Germany
\newline
$^{12}$Laboratoire Univers et Particules de Montpellier, Université de Montpellier,
Pl. Eug\`ne Bataillon - CC 72, 34095 Montpellier C\'edex 05, France
\newline
$^{13}$Department of Earth and Planetary Sciences, Weizmann Institute of Science,
234 Herzl Street, Rehovot 76100, Israel
\newline
$^{14}$Astronomisches Rechen-Institut, M\"onchhofstrasse 12-14 69120 Heidelberg Germany
\newline

$^\ast$Corresponding author E-mail: \texttt{g.anglada@qmul.ac.uk}
\newline
Authors listed in alphabetical order after corresponding author.
\newline

\textbf{At a distance of 1.295 parsecs \cite{vanleeuwen2007}, the red-dwarf
Proxima Centauri  ($\alpha$ Centauri C, GL 551, HIP 70890, or simply Proxima) is
the Sun's closest stellar neighbour and one of the best studied low-mass stars.
It has an effective temperature of only $\sim$ 3050 K, a luminosity of $\sim$0.1
per cent solar, a measured radius of 0.14 R$_\odot$ \cite{boyajian:2012} and a
mass of about 12 per cent the mass of the Sun. Although Proxima is considered a
moderately active star, its rotation period is $\sim$ 83 days \cite{kiraga2007},
and its quiescent activity levels and X-ray luminosity \cite{guedel:2004} are
comparable to the Sun's. New observations reveal the presence of a small planet
orbiting Proxima with a minimum mass of 1.3~Earth masses and an orbital period
of $\sim$11.2 days. Its orbital semi-major axis is $\sim0.05$ AU, with an
equilibrium temperature in the range where water could be liquid on its surface
\cite{hz:2013}.}

The results presented here consist of the analysis of previously obtained
Doppler measurements (pre-2016 data), and the confirmation of a signal in a
specifically designed follow-up campaign in 2016. The Doppler data comes from 
two precision radial velocity instruments, both at the European Southern
Observatory (ESO): the High Accuracy Radial velocity Planet Searcher (HARPS) and
the Ultraviolet and Visual Echelle Spectrograph (UVES). HARPS is a
high-resolution stabilized echelle spectrometer installed at the ESO 3.6m
telescope (La Silla observatory, Chile), and is calibrated in wavelength using
hollow cathode lamps. HARPS has demonstrated radial velocity
measurements at $\sim$1~\ms precision over time-scales of years
\cite{pepe:2011}, including on low-mass stars \cite{anglada:2012a}. All HARPS
spectra were extracted and calibrated with the standard ESO Data Reduction
Software, and radial velocities were measured using a least-squares template
matching technique \cite{anglada:2012a}. HARPS data is separated into two
datasets. The first set includes all data obtained before 2016 by several
programmes (HARPS pre-2016). The second HARPS set comes from the more recent
\emph{Pale Red Dot} campaign (PRD hereafter), which was designed to eliminate
period ambiguities using new HARPS observations and quasi-simultaneous
photometry. The HARPS~PRD observations consisted of obtaining one spectrum
almost every night between Jan 19th and March 31st 2016. The UVES observations
used the Iodine cell technique \cite{butler:1996} and were obtained in the
framework of the UVES survey for terrestrial planets around M-dwarfs between
2000 and 2008. The spectra were extracted using the standard procedures of the
UVES survey \cite{kuerster:2003}, and new radial velocities were re-obtained
using up-to-date Iodine reduction codes \cite{arriagada:2013}. Since systematic
calibration errors produce correlations among observations within each night
\cite{berdinas:2016}, we consolidated Doppler measurements through nightly
averages to present a simpler and more conservative signal search. This led to
72 UVES, 90 HARPS pre-2016 and 54 HARPS~PRD epochs. The PRD photometric
observations were obtained using the Astrograph for the South Hemisphere II
telescope (ASH2 hereafter \cite{ash2}, SII and H$_\alpha$ narrowband filters)
and the Las Cumbres Observatory Global Telescope network (LCOGT.net
\cite{lcogt}, Johnson B and V bands), over the same time interval and similar
sampling as the HARPS~PRD observations. Further details about each campaign and
the photometry are detailed in the methods section. All time-series used in this
work in the online version of the paper as Source data.

\begin{figure}[t]
\center
\includegraphics[angle=0, width=0.8\textwidth, clip]{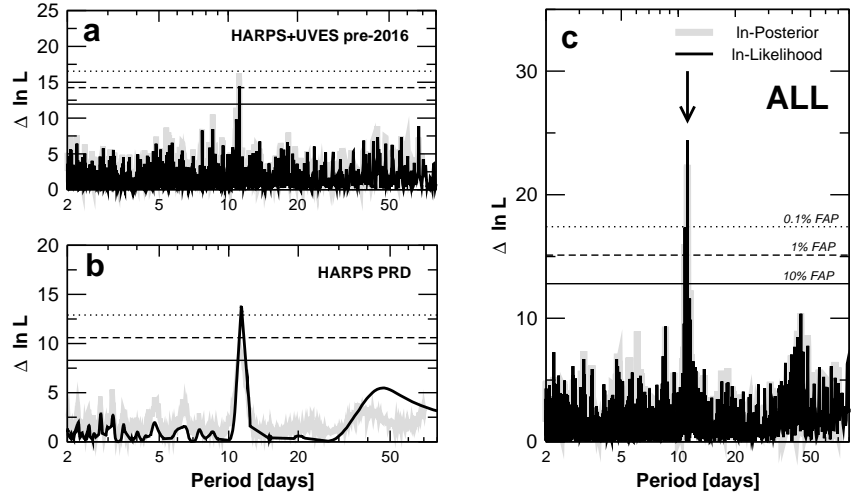}
\caption{\textbf{Detection of a Doppler signal at 11.2 days}. Detection
periodograms of the 11.2~day signal in the HARPS+UVES pre-2016 data (panel a),
and using the HARPS Pale Red Dot campaign only (panel b). Panel c contains the
periodogram obtained after combining all datasets. Black lines correspond to
the $\Delta \ln L$ statistic, while the gray thick represent the logarithm of the
Bayesian posterior density (see text, arbitrary vertical offset applied to for visual comparison of the two statistics). The horizontal solid,
dashed and dotted lines represent a 10, 1, and 0.1 per cent false alarm
probability thresholds of the frequentist analysis, respectively. }
\label{fig:periodograms}
\end{figure}

The search and significance assessment of signals were performed using
frequentist \cite{baluev:2013} and Bayesian \cite{tuomi:2014:uves} methods.
Periodograms in Figure~\ref{fig:periodograms} represent the improvement of some
reference statistic as a function of trial period, with the peaks representing
the most probable new signals. The improvement in the logarithm of the
likelihood function $\Delta \ln L$ is the reference statistic used in the
frequentist framework, and its value is then used to assess the false-alarm
probability (or FAP) of the detection \cite{baluev:2013}. A FAP below 1\% is
considered suggestive of periodic variability, and anything below 0.1\% is
considered to be a significant detection. In the Bayesian framework, signals are
first searched using a specialized sampling method \cite{haario:2006} that
enables exploration of multiple local maxima of the posterior density (the
result of this process are the gray lines in Figure~\ref{fig:periodograms}), and
significances are then assessed by obtaining the ratios of \emph{evidences} of
models. If the evidence ratio exceeds some threshold (e.g.
$B_{\rm 1}/B_{\rm 0} > 10^3$), then the model in the numerator (with one planet)
is favoured against the model in the denominator (no planet).

\begin{figure}[t]
\center
\includegraphics[angle=0, width=0.8\textwidth, clip]{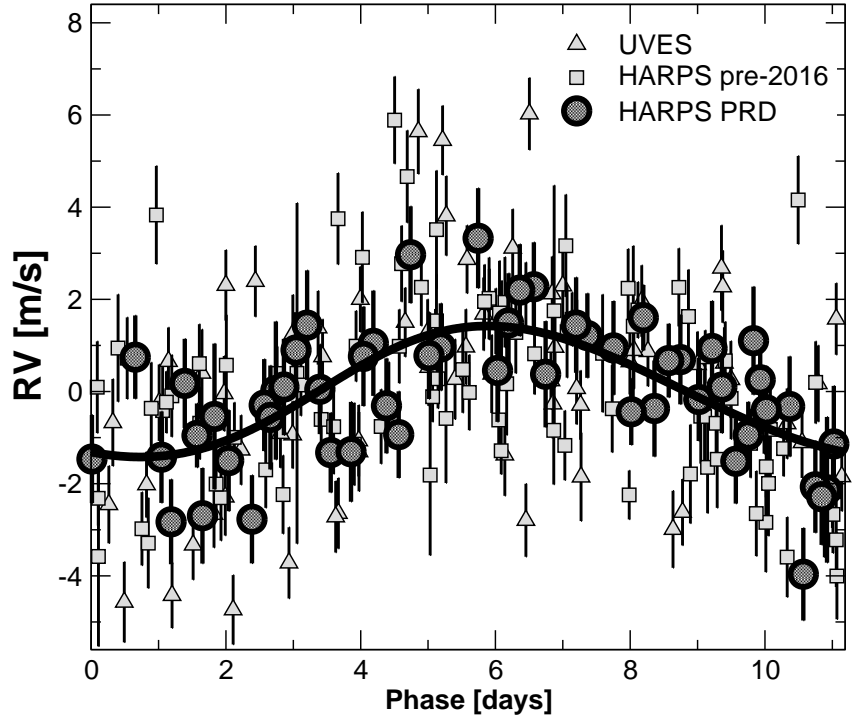}
\caption{\textbf{All datasets folded to the 11.2 days signal}. Radial velocity
measurements phase folded at the 11.2 day period of the planet candidate for 16
years of observations. Although its nature is unclear, a second signal at
P$\sim$ 200 days was fitted and subtracted from the data to produce this plot
and improve visualization. Circles correspond to HARPS~PRD, squares are
HARPS~pre-2016 and triangles are UVES observations. The black line represents 
the best Keplerian fit to this phase folded representation of the data. Error 
bars correspond to formal 1-$\sigma$ uncertainties.}\label{fig:phased}
\end{figure}

A well isolated peak at $\sim$11.2 days was recovered when analyzing all the
night averages in the pre-2016 datasets (Figure~\ref{fig:periodograms}, panel
a). Despite the significance of the signal, the analysis of pre-2016 subsets
produced slightly different periods depending on the noise assumptions and which
subsets were considered. Confirmation or refutation of this signal at 11.2 days
was the main driver for proposing the HARPS~PRD campaign. The analysis of the
HARPS~PRD data revealed a single significant signal at the same $\sim 11.3\pm
0.1$~day period  (Figure~\ref{fig:periodograms}, panel b), but period
coincidence alone did not prove consistency with the pre-2016 data. Final
confirmation was achieved when all the sets were combined
(Figure~\ref{fig:periodograms}, panel c). In this case statistical significance
of the signal at 11.2 days increases dramatically (false-alarm probability $<$
10$^{-7}$, Bayesian evidence ratio $B_{1,0} > 10^{6}$). This implies that not
only the period, but also the amplitude and phase are consistent during the 16
years of accumulated observations (see Figure~\ref{fig:phased}). All analyses
performed with and without correlated-noise models produced consistent results.
A second signal in the range of 60 to 500 days was also detected, but its nature
is still unclear due to stellar activity and inadequate sampling.

\begin{figure}[t]
\center
\includegraphics[angle=0, width=0.35\textwidth, clip]{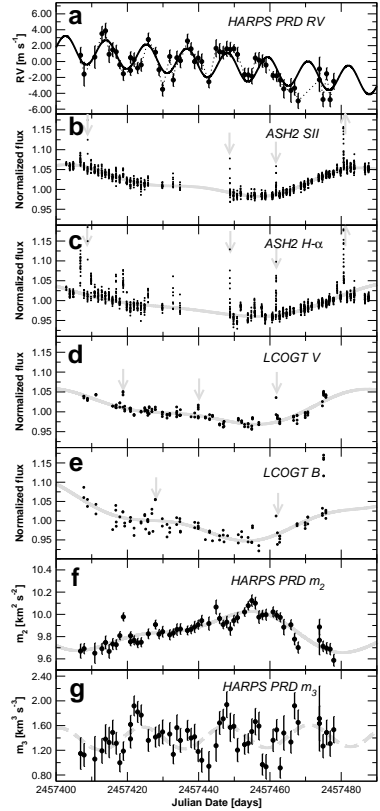}

\caption{\textbf{Time-series obtained during the Pale Red Dot campaign}. 
HARPS-PRD radial velocity measurements (panel a), quasi-simultaneous photometry
from ASH2 (panels b and c) and LCOGT (panels d and e) and central moments of the
mean line profiles (panels f and g). The solid lines show the best fits.  A
dashed line indicates a signal that is not sufficiently significant. Excluded
measurements likely affected activity events (e.g. flares) are marked with grey
arrows. The photometric time-series and $m_2$ all show evidence of the same
$\sim$80 day modulation. Error bars correspond to formal 1-$\sigma$
uncertainties.}\label{fig:prddata}

\end{figure}

Stellar variability can cause spurious Doppler signals that mimic planetary
candidates, especially when combined with uneven sampling \cite{rajpaul:2016,
kuerster:2003}. To address this, the time-series of the photometry and
spectroscopic activity indices were also searched for signals. After removing
occasional flares, all four photometric time-series show the same clear
modulation over $P\sim 80$ nights (panels b, c, d and e in
Figure~\ref{fig:prddata}), which is consistent with the previously reported
photometric period of $\sim$83~d \cite{kiraga2007}. Spectroscopic activity
indices were measured on all HARPS spectra, and their time-series were
investigated as well. The width of the spectral lines (measured as the variance
of the mean line, or $m_2$) follows a time-dependence almost identical to the
light curves, a behaviour that has already been reported for other M-dwarf stars
\cite{bonfils:2007}. The time-series of indices based on chromospheric emission
lines (e.g. H$_\alpha$) do not show evidence of periodic variability, even after
removing data points likely affected by flares. We also investigated possible
correlations of the Doppler measurements with activity indices by including
linear correlation terms in the Bayesian model of the Doppler data. While some
indices do show hints of correlation in some campaigns, including them in the
model produces lower probabilities due to overparameterization. Flares have very
little effect on our Doppler velocities, as has already been suggested by
previous observations of Proxima \cite{barnes:2014}. More details are provided
in the methods section and as Extended Data Figures. Since the analysis of the
activity data failed to identify any stellar activity feature likely to generate
a spurious Doppler signal at 11.2~days, we conclude that the variability in the
data is best explained by the presence of a planet (Proxima~b, hereafter)
orbiting the star. All available photometric light curves were searched for
evidence of transits, but no obvious transit-like features were detectable in
our light curves. We used Optimal Box-Least-Squares codes \cite{ofir:2014} to
search for candidate signals in data from the All Sky Automatic Survey
\cite{kiraga2007}. No significant transit signal was found down to a depth of
about 5\% either. The preferred orbital solution and the putative properties of
the planet and transits are given in Table \ref{tab:parameters}.

The Doppler semi-amplitude of Proxima~b ($\sim1.4$ \ms) is not particularly
small compared to other reported planet candidates \cite{pepe:2011}. The uneven
and sparse sampling combined with longer-term variability of the star seem to be
the reasons why the signal could not be unambiguously confirmed with pre-2016
rather than the amount of data accumulated. The corresponding minimum planet
mass is $\sim 1.3$ M$_\oplus$. With a semi-major axis of $\sim$0.05 AU, it lies
squarely in the center of the classical habitable zone for Proxima
\cite{hz:2013}. As mentioned earlier, the presence of another super-Earth mass
planet cannot yet be ruled out at longer orbital periods and Doppler
semi-amplitudes $<$3 \ms. By numerical integration of some putative orbits, we
verified that the presence of such an additional  planet would not compromise
the orbital stability of Proxima~b.

Habitability of planets like Proxima~b -in the sense of sustaining an atmosphere
and liquid water on its surface- is a matter of intense debate. The most common
arguments against habitability are tidal locking, strong stellar magnetic field,
strong flares, and high UV \& X-ray fluxes; but none of these have been proven
definitive. Tidal locking does not preclude a stable atmosphere via global
atmospheric circulation and heat redistribution\cite{kopparapu:2016}. The average
global magnetic flux density of Proxima is 600$\pm$150 Gauss \cite{reiners:2008},
which is quite large compared to the Sun's value of 1~G. However, several studies
have shown that planetary magnetic fields in tidally locked planets can be strong
enough to prevent atmospheric erosion by stellar magnetic
fields\cite{vidotto:2013} and flares, \cite{zuluaga:2013}. Because of its close-in
orbit, Proxima~b suffers X-ray fluxes $\sim$400 times that of Earth's, but studies
of similar systems indicate that atmospheric losses can be relatively small
\cite{bolmont:2016}. Further characterization of such planets can also inform us
about the origin and evolution of terrestrial planets. For example, forming
Proxima~b from in-situ disk material is implausible because disk models for small
stars would contain less than 1 $M_{\rm Earth}$ of solids within the central AU.
Instead, either 1) the planet migrated in via type I migration \cite{tanaka:2002},
2) planetary embryos migrated in and coalesced at the current planet's orbit, or
3) pebbles/small planetesimals migrated via aerodynamic drag \cite{weid:1977} and
later coagulated into a larger body. While migrated planets and embryos
originating beyond the ice-line would be volatile rich,  pebble migration would
produce much drier worlds. In this sense, a warm terrestrial  planet around
Proxima offers unique follow-up opportunities to attempt further characterization
via transits -on going searches-, via direct imaging and high-resolution
spectroscopy in the next decades \cite{snellen:2015}, and --maybe-- robotic
exploration in the coming centuries \cite{lubin:2016}.

\noindent

\textbf{Acknowledgements.} 

We thank Enrico Gerlach, Rachel Street and Ulf Seemann for their
support to the science preparations. We thank Javier Pascual-Granado and
Rafael Garrido for their insights on the properties of the putative
signals in the Doppler time-series. 

We thank Predrag Micakovic, Matthew M.
Mutter (QMUL), Rob Ivison, Gaitee Hussain, Ivo Saviane, Oana Sandu, Lars
Lindberg Christensen, Richard Hook and the personnel at La Silla (ESO) for
making the Pale Red Dot campaign possible.
The authors acknowledge support from funding grants;
Leverhulme Trust/UK RPG-2014-281 (HRAJ, GAE, MT), 
MINECO/Spain AYA-2014-54348-C3-1-R (PJA, CRL, ZMB, ER), 
MINECO/Spain ESP2014-54362-P (MJLG),
MINECO/Spain AYA-2014-56637-C2-1-P(JLO, NM),
J.A./Spain 2012-FQM1776 (JLO, NM),
CATA-Basal/Chile PB06 Conicyt (JSJ), 
Fondecyt/Chile project \#1161218 (JSJ), 
STFC/UK ST/M001008/1 (RPN, GALC, GAE), 
STFC/UK ST/L000776/1 (JB), 
ERC/EU Starting Grant \#279347 (AR, LS, SVJ), 
DFG/Germany Research Grants RE 1664/9-2(AR), 
RE 1664/12-1(MZ), 
DFG/Germany Colloborative Research Center 963(CJM, SD),
DFG/Germany Research Training Group 1351(LS), and 
NSF/USA grant AST-1313075 (ME).
Based on observations made with ESO Telescopes at the La Silla Paranal 
Observatory under programmes 096.C-0082 and 191.C-0505. Observations were 
obtained with ASH2, which is supported by the
Instituto de Astrof\'isica de Andaluc\'ia and Astroimagen company.
This work makes use of observations from the LCOGT network.
We acknowledge the effort of
the UVES/M-dwarf and the HARPS/Geneva teams, which
obtained a substantial amount of the data used in this work.

\noindent
\textbf{Author contributions.}
GAE led the PRD campaign, observing proposals and organized the manuscript.
PJA led observing proposals, and organized and supported the IAA team through research grants.
MT obtained the early signal detections and most Bayesian analyses.
JSJ, JB, ZMB and HRAJ participated in the analyses and obtained activity measurements. Zaira M. Berdi\~nas also led observing proposals. 
HRAJ funded several co-authors via research grants.
MK and ME provided the extracted UVES spectra, 
and RPB re-derived new RV measurements.
CRL coordinated photometric follow-up campaigns.
ER led the ASH2 team and related reductions (MJLG, IC, JLO, NM).
YT led the LCOGT proposals, campaign and reductions.
MZ obtained observations and performed analyses on HARPS and UVES spectra.
AO analysed time-series and transit searches.
JM, SVJ and AR analyzed stellar activity data. AR funded several 
co-authors via research grants.
RPN, GALC, SJP, SD \& BG did dynamical and studied the planet formation context.
MK provided early access to time-series from the ASAS survey.
CJM and LFS participated in the HARPS campaigns.
All authors contributed to the preparation of observing proposals and the manuscript.
\\

Reprints and permissions information is available at
\texttt{www.nature.com/reprints}.
\\

The authors declare that they do not have any competing financial interests.
\\

Correspondence and requests for materials should be addressed to Guillem
Anglada-Escud\'e, \texttt{g.anglada@qmul.ac.uk}

\newpage

\begin{table}[h]
\begin{minipage}{\textwidth}
\caption{Stellar properties, Keplerian parameters, and derived quantities. The
estimates are the maximum \emph{a posteriori} estimates and the uncertainties
are expressed as 68\% credibility intervals. We only provide an upper limit for
the eccentricity (95\% confidence level). Extended Data 
Table\ref{tab:allpars} contains the list of all the model parameters.
}\label{tab:parameters}
\begin{center}
%\small
\begin{tabular}{lll}
\hline \hline
Stellar properties & Value & Reference\\
\hline
Spectral type                              & M5.5V                       & \cite{boyajian:2012}  \\
Mass$_*$/Mass$_{\rm Sun}$                  & 0.120 [0.105,0.135]         & \cite{delfosse:2000}  \\
Radius$_*$/R$_{\rm Sun}$                   & 0.141 [0.120,0.162]         & \cite{boyajian:2012}  \\
Luminosity$_*$/ L$_{\rm Sun}$              & 0.00155  [0.00149, 0.00161] & \cite{boyajian:2012}  \\
Effective temperature [K]                  & 3050  [2950, 3150]          & \cite{boyajian:2012}  \\
Rotation period [days]                     & $\sim$ 83                   & \cite{kiraga2007}     \\
Habitable zone range [AU]                  & $\sim$ 0.0423 -- 0.0816     & \cite{kopparapu:2016} \\
Habitable zone periods [days]              & $\sim$ 9.1--24.5            & \cite{kopparapu:2016} \\
\\
Keplerian fit & Proxima~b                     \\
\hline
Period [days]          & 11.186 [11.184, 11.187] \\
Doppler amplitude [ms$^{-1}$] & 1.38 [1.17, 1.59]              \\
Eccentricity [-]                & $<$0.35                      \\
Mean longitude $\lambda=\omega+M_0$ [deg] & 110 [102, 118]  \\
Argument of periastron $w_{0}$ [deg]   & 310 [0,360]                        \\
\\
Statistics summary\\
\hline
Frequentist false alarm probability & $ 7 \times 10^{-8}$ \\
Bayesian odds in favour B$_{\rm 1}$/B$_{\rm 0}$   & $ 2.1\times 10^{7}$ \\
UVES Jitter [\ms] & 1.69 [1.22, 2.33] \\
HARPS pre-2016 Jitter [\ms] & 1.76 [1.22, 2.36] \\
HARPS PRD Jitter [\ms] & 1.14 [0.57, 1.84]\\
\\
Derived quantities\\
\hline
Orbital semi-major axis $a$ [AU]   & 0.0485 [0.0434, 0.0526] \\
Minimum mass $m_{p} \sin i$ [M$_{\oplus}$]         & 1.27 [1.10, 1.46]           \\
Eq. black body temperature [K] &  234 [220, 240]             \\
Irradiance compared to Earth's &   65\%              \\
Geometric probability of transit  &  $\sim$1.5\%   \\
Transit depth (Earth-like density)  &  $\sim$0.5\%   \\
\hline \hline
\end{tabular}
\end{center}
\end{minipage}
\end{table}

\newpage

\section*{Methods}
\renewcommand{\figurename}{\textbf{Extended Data Figure}}
\renewcommand{\tablename}{\textbf{Extended Data Table}}
\setcounter{figure}{0}
\setcounter{table}{0}

\section{Statistical frameworks and tools}\label{sec:statistics}

The analyses of time-series including radial velocities and activity indices
were performed by frequentist and Bayesian methods. In all cases,
significances were assessed using model comparisons by performing global multi-parametric fits to the data. Here we provide a minimal overview of the methods and assumptions used throughout the paper.

\subsection{Bayesian statistical analyses.}\label{sec:bayesain}

The analyses of the radial velocity data were performed by applying posterior
sampling algorithms called Markov chain Monte Carlo (MCMC) methods. We used the
adaptive Metropolis algorithm\cite{haario:2001} that has previously been applied
to such radial velocity data sets \cite{tuomi:2013b,tuomi:2014:uves}. This
algorithm is simply a generalised version of the common Metropolis-Hastings
algorithm \cite{metropolis:1953,hastings:1970} that adapts to the posterior
density based on the previous members of the chain.

Likelihood functions and posterior densities of models with periodic signals are
highly multimodal (i.e. peaks in periodograms). For this reason, in our Bayesian
signal searches we applied the delayed rejection adaptive Metropolis (DRAM)
method \cite{haario:2006}, that enables efficient jumping of the chain between
multiple modes by postponing the rejection of a proposed parameter vector by
first attempting to find a better value in its vicinity. For every given model,
we performed several posterior samplings with different initial values to ensure
convergence to a unique solution. When we identified two or more significant
maxima in the posterior, we typically performed several additional samplings
with initial states close to those maxima. This enabled us to evaluate all of
their relative significances in a consistent manner. We estimated the marginal
likelihoods and the corresponding Bayesian \emph{evidence ratios} of different
models by using a simple method\cite{newton:1994}. A more detailed description
of these methods can be found in elsewhere\cite{tuomi:2014b}.

\subsection{Statistical models : Doppler model and likelihood function.}
\label{sec:likelihood}

Assuming radial velocity measurements $m_{i,{\rm INS}}$ at some instant $t_i$
and instrument INS, the likelihood function of the observations (probability of
the data given a model) is given by
\begin{eqnarray}\label{eq:likelihood}
  L &=& \prod_{\rm INS} \prod_i l_{i,{\rm INS}}\, \\
  l_{i,{\rm INS}} &=&
  \frac{1}{\sqrt{2\pi \left( \sigma_i^2 + \sigma_{\rm INS}^{2}\right)}}
  \exp \Bigg\{ - \frac{1}{2} \,\,
  \frac{\epsilon_{i,{\rm INS}}^{2}}{\sigma_i^2 + \sigma_{\rm INS}^{2}} \Bigg\} ,\\
  \epsilon_{i,{\rm INS}} &=&
  m_{i,{\rm INS}} - \Bigg\{
      \gamma_{\rm INS} +
      \dot{\gamma} \Delta t_{i} +
            \kappa(\Delta t_{i}) +
      {\rm MA}_{i,{\rm INS}} +
      {\rm A}_{i,{\rm INS}}
  \Bigg\}    \,  ,
      \\
  \Delta t_i &=& t_i-t_0 \, \,
\end{eqnarray}
\noindent where $t_0$ is some reference epoch. This reference epoch
can be arbitrarily chosen, often as the beginning of the
time-series or a mid-point of the observing campaigns. The other terms are:

\begin{itemize}

\item $\epsilon_{i,{\rm lNS}}$ are the residuals to a fit. We assume that each
$\epsilon_{i,{\rm lNS}}$ is a Gaussian random variable with a zero mean and a
variance of  $\sigma_{i}^{2} + \sigma_{\rm INS}^{2}$, where $\sigma_{i}^{2}$ is
the reported uncertainty of the $i$-th measurement and $\sigma_{\rm INS}^{2}$ is
the \emph{jitter parameter} and represents the excess white noise not included
in $\sigma_{i}^{2}$.

\item $\gamma_{\rm INS}$ is the \emph{zero-point velocity} of each instrument.
Each INS can have a different zero-point depending on how the radial velocities
are measured and how the wavelengths are calibrated.

\item $\dot{\gamma}$ is a \emph{linear trend parameter} caused by a long term
acceleration.

\item The term $\kappa(\Delta t_i)$ is the superposition of $k$ Keplerian
signals evaluated at $\Delta t_{i}$. Each Keplerian signal depends on five
parameters: the \emph{orbital period} $P_p$, \emph{semi-amplitude} of
the signal $K_p$, mean anomaly $M_{0,p}$, which represents the phase of the
orbit with respect to the periastron of the orbit at $t_0$, \emph{orbital eccentricity} $e_p$
that goes from $0$ (circular orbit) to $1$ (unbound parabolic orbit), and the
\emph{argument of periastron} $\omega_p$, which is the angle on the orbital
plane with respect to the plane of the sky at which the star goes through the
periastron of its orbit (the planet's periastron is at $\omega_p+180$ deg).
Detailed definitions of the parameters can be found elsewhere\cite{wright:2009}.
\item  The Moving Average term
\begin{eqnarray}
 {\rm MA}_{i, {\rm INS}} =
 \phi_{\rm INS}
 \exp \Bigg\{ \frac{t_{i-1} - t_{i}}{\tau_{\rm INS}} \Bigg\}
 \epsilon_{i-1,{\rm INS}}
\end{eqnarray}
\noindent is a simple parameterization of possible correlated noise that depends
on the residual of the previous measurement $\epsilon_{i-1,{\rm INS}}$. As for
the other parameters related to noise in our model, we assume that the
parameters of the MA function depend on the instrument; for example the
different wavelength ranges used will cause different properties of the instrumental
systematic noise. Keplerian and other physical processes also introduce
correlations into the data, therefore some degree of degeneracy between the MA
terms and the signals of interest is expected. As a result, including a MA term
always produces more conservative significance estimates than a model with
uncorrelated random noise only. The MA model is implemented through a
coefficient $\phi_{\rm INS}$ and a time-scale $\tau_{\rm INS}$. $\phi_{\rm INS}$
quantifies the strength of the correlation between the $i$ and $i-1$ measurements. It is
bound between $-1$ and $1$ to guarantee that the process is stationary (i.e. the
contribution of the MA term does not arbitrarily grow over time). The
exponential smoothing is used to decrease the strength of the correlation
exponentially as the difference $t_{i} - t_{i-1}$ increases\cite{scargle:1981}.

\item Linear correlations with activity indices can also be included in the
model in the following manner,
\begin{eqnarray}
{\rm A}_{i,{\rm INS}} = \sum_\xi C_{\xi,{\rm INS}}
\, \, \xi_{i,{\rm INS}}
\end{eqnarray}
\noindent where $\xi$ runs over all the activity indices used to model each
INS dataset (e.g. $m_2$, $m_3$, S-index, etc. whose description is provided below). To
avoid any confusion with other discussions about correlations, we call these
$C_{\xi, {\rm INS}}$ \emph{activity coefficients}. Note that each activity
coefficient $C_{\xi, {\rm INS}}$ is associated to one activity index ($\xi_i$)
obtained simultaneously with the i-th radial velocity measurement (e.g.
chromospheric emission from the H$_\alpha$ line, second moment of the mean-line
profile, interpolated photometric flux, etc.). When fitting a model to the data,
an activity coefficient significantly different from $0$ indicates evidence of
Doppler variability correlated with the corresponding activity index. Formally
speaking, these $C_{\xi , {\rm INS}}$ correspond to the coefficient of the first
order Taylor expansion of a physical model for the apparent radial velocities as
a function of the activity indices and other physical properties of the star.

\end{itemize}

A simplified version of the same likelihood model is used when analyzing
time-series of activity indices. That is, when searching for periodicities in
series other than Doppler measurements, the model will consist of the
$\gamma_{\rm INS}$ zero-points, a linear trend term $\dot{\gamma}\Delta t_i$, and a
sum of $n$ sinusoids
\begin{eqnarray}
\hat\kappa(t_i, \vec{\theta}) &=&
\sum^n_k
\left(A_k \sin \frac{2\pi \Delta t_i}{P_k} +
B_k \cos \frac{2\pi \Delta t_i}{P_k}\right)\label{eq:circular}
\end{eqnarray}
\noindent where each $k$-th sinusoid has three parameters $A_k$, $B_k$, and
$P_k$ instead of the five Keplerian ones. Except for the period parameters and
the jitter terms, this model is linear with all the other parameters, which
allows a relatively quick computation of the likelihood-ratio periodograms.

\subsection{Bayesian prior choices.}\label{sec:priors}

As in any Bayesian analysis, the prior densities of the model parameters have to
be selected in a suitable manner (for example see \cite{tuomi:2012}). We used
uniform and uninformative distributions for most of the parameters apart from a
few, possibly significant, exceptions. First, as we used a parameter $l = \ln P$
in the MCMC samplings instead of the period $P$ directly, the uniform prior
density $\pi(l) = c$ for all $l \in [\ln T_{0}, \ln T_{\rm max}]$, where $T_{0}$
and $T_{\rm max}$ are some minimum and maximum periods, does not correspond to a
uniform prior in $P$. Instead, this prior corresponds to a period prior such
that $\pi(P) \propto P^{-1}$ \cite{tuomi:2013:gj163}. We made this choice
because the period can be considered a ``scale parameter'' for which an
uninformative prior is one that is uniform in $\ln P$ \cite{berger:1980}. We
selected the parameter space of the period such that $T_{0} = 1$ day and $T_{\rm
max} = T_{\rm obs}$, where $T_{\rm obs}$ is the baseline of the combined data.

For the semi amplitude parameter $K$, we used a $\pi(K) = c$ for all $K \in [0,
K_{\rm max}]$, where $K_{\rm max}$ was selected as $K_{\rm max} = 10$ ms$^{-1}$
because the RMSs of the Doppler series did not exceed 3 ms$^{-1}$ in any of the
sets. Following previous works \cite{tuomi:2013:gj163,anglada:2013}, we chose
the prior for the orbital eccentricities as $\pi(e) \propto \mathcal{N}(0,
\Sigma_{e}^{2})$, where $e$ is bound between zero (circular orbit) and 1. We set
this $\Sigma_{e}^{2} = 0.3$ to penalize high eccentricities while keeping the
option of high $e$ if the data strongly favours it.

We also used an informative prior for the excess white noise parameter of
$\sigma_{INS}$ for each instrument. Based on analyses of a sample of M dwarfs
\cite{tuomi:2014:uves}, this ``stellar jitter'' is typically very close to a
value of 1 ms$^{-1}$. Thus, we used a prior such that $\pi(\sigma_{l}) \propto
\mathcal{N}(\mu_{\sigma}, \sigma_{\sigma}^{2})$ such that the parameters were
selected as $\mu_{\sigma} = \sigma_{\sigma} =$ 1 ms$^{-1}$. Uniform priors were
used in all the activity coefficients $C_\xi \in [-C_{\xi,max}, C_{\xi,max}]$.
For practical purposes, the time-series of all activity indices were mean
subtracted and normalized to their RMS. This choice allows us to select the
bounds of the activity coefficients for the renormalized time-series as
$\hat{C}_{\xi,max} =$ 3 ms$^{-1}$, so that adding correlation terms does not
dramatically increase the RMS of the Doppler time-series over the initially
measured RMS of $<3$ ms$^{-1}$ (same argument as for the prior on $K$). This
renormalization is automatically applied by our codes at initialization.

\subsection{Search for periodicities and significances in a frequentist
framework.}\label{sec:frequentist}

Periodograms are plots representing a figure-of-merit derived from a fit against
the period of a newly proposed signal. In the case of unevenly sampled data, a
very popular periodogram is the Lomb-Scargle periodogram (or LS)
\cite{lomb:1976, scargle:1982} and its variants like the Floating-mean
periodogram\cite{zechmeister:2009} or the F-ratio
periodogram\cite{cumming:2004}. In this work we use likelihood ratio
periodograms, which represent the improvement of the likelihood statistic when
adding a new sinusoidal signal to the model. Due to intrinsic non-linearities in
the Keplerian/RV modelling, optimizing the likelihood statistic is more
computationally intensive than the classic LS-like periodograms
\cite{ferraz_mello:1981, zechmeister:2009}). On the other hand the likelihood
function is a more general and well-behaved statistic which, for example, allows
for the optimisation of the noise parameters (e.g.~\textit{jitter}, and fit
correlated noise models at the signal search level). Once the maximum likelihood
of a model with one additional planet is found (highest peak in the
periodogram), its false-alarm probability can then be easily computed
\cite{baluev:2009, baluev:2012}. In general, a false-alarm probability
of 1\% is needed to claim hints of variability, and a value below 0.1$\%$ is
considered necessary to claim a significant detection.

\section{Spectroscopic datasets}\label{sec:spectraldata}

\subsection{New reduction of the UVES M-dwarf programme data.}
\label{sec:uves}

Between 2000 and 2008, Proxima was observed in the framework of a precision RV
survey of M dwarfs in search for extrasolar planets with the Ultraviolet and
Visual Echelle Spectrograph (UVES) installed in the Very Large Telescope (VLT)
unit 2 (UT2). To attain high-precision RV measurements, UVES was self-calibrated
with its iodine gas absorption cell operated at a temperature of $70^\circ$ C. 
The image slicer $\# 3$ was chosen which redistributes the light from a
$1^{\prime \prime} \times 1^{\prime \prime }$ aperture along the chosen
$0.3^{\prime \prime }$ wide slit. In this way, a resolving power of $R=100,000 -
120,000$ was attained. At the selected central wavelength of $600~\mathrm{nm}$,
the useful spectral range containing iodine ($I_2$) absorption lines ($\approx
500 - 600~\mathrm{nm}$) falls entirely on the better quality detector of the
mosaic of two $4K\times 2$K CCDs. More details can be found in the several
papers from the UVES survey\cite{kuerster:2003, endl:2008, zechmeister:2009}.

The extracted UVES spectra include 241 observations taken through the Iodine
cell, three template (no Iodine) shots of Proxima, and three spectra of the
rapidly rotating B star HR 5987 taken through the Iodine cell as well, and almost
consecutive to the three template shots. The B star has a smooth spectrum devoid
of spectral features and it was used to calibrate the three template observations
of the target. Ten of the Iodine observations of Proxima were eliminated due to low
exposure levels. The remaining 231 iodine shots of Proxima were taken on 77
nights, typically 3 consecutive shots per night.

The first steps in the process of $I_2$ calibrated data consists of constructing
the high signal to noise template spectrum of the star without iodine: 1) a
custom model of the UVES instrumental profile is generated based on the
observations of the B star by forward modeling the observations using a
higher-resolution ($R=700,000 - 1,000,000$) template spectrum of the $I_2$ cell
obtained with the McMath Fourier Transform Spectrometer (FTS) on Kitt Peak, 2)
the three template observations of Proxima are then co-added and filtered for
outliers, and 3) based on the instrument profile model and wavelength solution
derived from the three B star observations, the template is deconvolved with our
standard software \cite{arriagada:2013}. After the creation of the stellar
template, the 231 iodine observations of Proxima were then run through our
standard precision velocity code \cite{butler:1996}. The resulting standard
deviation of the 231 un-binned observations is 2.58 ms$^{-1}$, and the standard
deviation of the 77 nightly binned observations is 2.30 ms$^{-1}$, which already
suggests an improvement compared to the 3.11 ms$^{-1}$ reported in the original
UVES survey reports\cite{endl:2008}. All the UVES spectra (raw) are publicly
available in their reduced form via ESO's archive at
\texttt{http://archive.eso.org/cms.html}. Extracted spectra are not produced for
this mode of UVES operation, but they are available upon request.

\subsection{HARPS GTO.}\label{sec:harpsgto}
The initial HARPS-Guaranteed Time Observations programme was led by
Michel Mayor (ESO ID : 072.C-0488). 19 spectra were obtained between May 2005
and July 2008. The typical integration time ranges between 450 and 900~s.

\subsection{HARPS M-dwarfs.} \label{sec:harpsmdwarf}
Led by X. Bonfils and collaborators, it consists of ESO programmes 082.C-0718
and 183.C-0437. It produced 8 and 46 measurements respectively with integration
times of 900~s in almost all cases\cite{bonfils:2013}.

\subsection{HARPS high-cadence.} \label{sec:harpsctb}
This program consisted of two 10 night runs (May 2013, and Dec 2013, ESO ID:
191.C-0505) and was led and executed by several authors of this paper. Proxima
was observed on two runs 
\begin{itemize}
\item May 2013 - 143 spectra obtained in three consecutive nights between May 4th and May 7th and 25 additional spectra between May 7th and May 16th with exposure times of 900~s.
\item Dec 2013 -23 spectra obtained between Dec 30th and
Jan 10th 2014 also with 900~s exposure times.
\end{itemize}

\noindent For simplicity in the presentation of the data and analyses, all HARPS
data obtained prior to 2016 (HARPS GTO, HARPS M-dwarfs, and HARPS high-cadence)
are integrated in the so-called HARPS pre-2016 set. The long-term Doppler
variability and sparse sampling makes the detection of the Doppler signal more
challenging in such a consolidated set than, for example, separating it into
subsets of contiguous nights. The latter strategy, however, necessarily requires
more parameters (offsets, jitter terms, correlated noise parameters) and
arbitrary choices on the sets to be used, producing strong degeneracies and
aliasing ambiguities in the determination of the favoured solution (11.2-d was
typically favoured, but alternative periods caused by a non-trivial window
function at 13.6-d, 18.3-d were also found to be possible). The data taken in
2016 exclusively corresponds to the new campaign specifically designed to
address the sampling issues.

\subsection{HARPS : Pale Red Dot campaign.} \label{sec:harpsprd}

PRD was executed between Jan 18th and March 30th, 2016. Few nights interruptions
were anticipated to allow for technical work and other time-critical
observations with HARPS. Of the 60 scheduled epochs, we obtained 56 spectra in
54 nights (two spectra were obtained in two of those nights). Integration times
were set to 1200~s, and observations were always obtained at the very end of
each night. All the HARPS spectra (raw, extracted and calibrated frames) are
publicly available in their reduced form via ESO's archive at
\texttt{http://archive.eso.org/cms.html}.

\section{Spectroscopic indices} \label{sec:indices}

Stellar activity can be traced by features in the stellar spectrum. For
example, changes in the line-profile shapes (symmetry and width) have been
associated to spurious Doppler shifts \cite{queloz:2001, bonfils:2007}.
Chromospheric emission lines are tracers of spurious Doppler variability in
the Sun and they are expected to behave similarly for other stars
\cite{robertson:2014}. We describe here the indices measured and used in our
analyses.

\subsection{Measurements of the mean spectral line profiles.} \label{sec:lines}
The HARPS Data Reduction Software provides two measurements of the mean-line
profile shapes derived from the cross-correlation function (CCF) of the stellar
spectrum with a binary mask. These are called the bisector span (or BIS) and
full-width-at-half-maximum (or FWHM) of the CCF \cite{bonfils:2013}. For
very late type stars like Proxima, all spectral lines are blended producing a
non-trivial shape of the CCF, and thus the interpretation of the usual
line-shape measurements is not nearly as reliable as in earlier type stars.
We applied the Least-Squares Deconvolution (LSD) technique \cite{donati:1997}
to obtain a more accurate estimate of the spectral mean line profile. This
profile is generated from the convolution of a kernel, which is a model
spectrum of line positions and intensities, with the observed spectrum. A
description of our implementation of the procedure, applied specifically to
crowded M-dwarf spectra is described in \cite{barnes:2012}. The LSD profile
can be interpreted as a probability function distribution that can then be
characterized by its central moments \cite{numerical}. We computed the second
($m_2$) and third ($m_3$) central moments of each LSD-profile of each
observation. More details of these indices and how they compare to other
standard HARPS cross-correlation measurements can be found in
\cite{berdinas:2016}. To eliminate the correlation of the profile moments
with the slope of the spectral energy distribution \cite{berdinas:2016}, we
corrected the SED and blaze function to match the same spectral energy distribution
of the highest S/N observation obtained with HARPS. Uncertainties were
obtained using an empirical procedure as follows: we derived all the $m_2$
and $m_3$ measurements of the high-cadence night of May 7th 2013 and fitted a
polynomial to each time-series. The standard deviation of the residuals to
that fit was then assumed to be the expected uncertainty for a S/N$\sim$20
(at reference echelle aperture number 60), which was the typical value for
that night's observations.  All other errors were then obtained by scaling
this standard deviation by a factor of $\frac{20}{S/N_{\rm obs}}$ for each
observation.

\subsection{Chromospheric indices.}\label{sec:chrom}
Chromospheric emission lines are tracers of spurious Doppler variability in
the Sun and they are expected to behave similarly for other stars
\cite{robertson:2014}. We describe here the indices computed and used in our
analyses.

\subsection{Chromospheric CaII H+K S-index.} \label{sec:sindex}
We calculated the CaII H+K fluxes following standard procedures 
\cite{jenkins06,jenkins08}, both the PRD data and the pre-2016 data were treated
the same. Uncertainties were calculated from the quadrature sum of the variance
in the data used within each bandpass.

\subsection{Chromospheric H$_\alpha$ emission.} \label{sec:halpha}
This index was measured in a similar way to the $S$-indices, such that we summed
the fluxes in the center of the lines, calculated to be 6562.808~\AA, this time
utilising square bandpasses of 0.678~\AA\ not triangular shapes, and those were
normalized to the summed fluxes of two square continuum band regions surrounding
each of the lines in the time series.  The continuum square bandpasses were
centered at 6550.870~\AA\ and 6580.309~\AA\ and had widths of 10.75~\AA\ and
8.75~\AA, respectively.  Again the uncertainties were calculated from the
quadrature sum of the variance of the data within the bandpasses.

\section{Photometric datasets} \label{sec:photometry}
\subsection{Astrograph Southern Hemisphere II.} \label{sec:ash2}

The ASH2 (Astrograph for the South Hemisphere II) telescope is a 40~cm robotic
telescope with a CCD camera STL11000 2.7K x 4K, and a field-of-view (FOV) of 54
x 82~arcmin. Observations were obtained in two narrow-band filters centered on
H$_\alpha$ and SII lines, respectively (H$_\alpha$ is centred on 656~nm, SII is
centered on 672~nm, and both filters have a Gaussian-like transmission with a
FWHM of 12~nm). The telescope is at SPACEOBS (San Pedro de Atacama Celestial
Explorations Observatory), at 2450~m above sea  level, located in the northern
Atacama Desert, in Chile. This telescope is managed and supported by the
Instituto de Astrof\'isica de Andaluc\'ia (Spain). During the present work, only
subframes with 40\% of the total field of view  were used, resulting in a useful
FOV of 21.6 $\times$ 32.8~arcmin. Approximately 20 images in each band of 100~s
of exposure time were obtained per night. In total, 66 epochs of about 100~min
each were obtained during this campaign. The number of images collected per
night was increased during the second part of the campaign (until about 40
images in each filter per night).

All CCD measurements were obtained by the method of synthetic aperture
photometry using a 2 $\times$ 2 binning. Each CCD frame was corrected in a
standard way for dark and flat-fielding. Different aperture sizes were also
tested in order to choose the best one for our observations. A number of nearby
and relatively bright stars within the frames were selected as check stars in
order to choose the best ones to be used as comparison stars. After checking
their stability, C2=HD\,126625 and C8=TYC\,9010-3029-1, were selected as main
comparison stars.

The basic photometric data were computed as magnitude differences in SII and
H$_\alpha$ filters for Var-X and  C2-X, with Var=Prox Cen and X=(C2+C8)/2.
Typical uncertainties of each individual data point are about 6.0~mmag, for both
SII and H$_\alpha$ filters. This usually leads to error-bars of about 1.3~mmag
in the determination of the mean levels of each epoch, assuming 20 points per
night once occasional strong activity episodes (such as flares) are removed for
the analysis of periodicities. For the analyses, these magnitudes were
transformed to relative flux measurements normalized to the mean flux over the
campaign.

\subsection{Las Cumbres Observatory Global Telescope network.}\label{sec:lcogt}

The Las Cumbres Observatory (LCOGT) is an organization dedicated to time-domain
astronomy \cite{lcogt}. To facilitate this, LCOGT operates a homogeneous network
of 1~m and 2~m telescopes on multiple sites around the world. The telescopes are
controlled by a single robotic scheduler, capable of orchestrating complex
responsive observing programs, using the entire network to provide uninterrupted
observations of any astronomical target of interest. Each site hosts between one
to three telescopes, which are configured for imaging and spectroscopy. The
telescopes are equipped with identical instruments and filters, which allows for
'network redundancy'. This means that observations can be seamlessly shifted to
alternate sites at any time if the scientific program requires it, or in the
event of poor weather.

Observations for the PRD campaign were obtained on the 1~m network every 24
hours in the B and V bands with the Sinistro (4K x 4K Fairchild CCD486) cameras,
which have a pixel scale of 0.38~arcsec and a FOV of 27 x 27 arcminutes. In
addition, B and V observations were taken every 12 hours with the SBIG (4K x 4K
Kodak KAF-6303E CCD) cameras, with a pixel scale of 0.46~arcsec and a FOV of 16
x 16 arcminutes. Exposure times ranged between 15 and 40~s and a total of 488
photometrically useful images were obtained during the campaign.

The photometric measurements were performed using aperture photometry with
AstroImageJ\cite{collins:2016} and DEFOT\cite{southworth:2014}. The aperture
sizes were optimized during the analysis with the aim of minimizing measurement
noise. Proxima Centauri and two non-variable comparison stars were identified in
a reference image and used to construct the detrended light curves. As with the
ASH2 curves, the LCOGT differential magnitudes were transformed to normalized
flux to facilitate interpretation and later analyses (see 
Figure~\ref{fig:prddata} in main article).

\section{Signals in time-series}\label{sec:signalsearches}

In this section we present a homogeneous analysis of all the time-series
(Doppler, activity and photometric ones) presented in this article. In all
periodograms, the black curve represents the search for a first signal. If one
first signal is identified, then a red curve represents the search for a second
signal. In the few cases where a second signal is detected, a blue curve
represents the search for a third signal. The period of Proxima~b is marked
with a green vertical line.

\subsection{Module of the Window function.} \label{sec:window}

\begin{figure}[t]
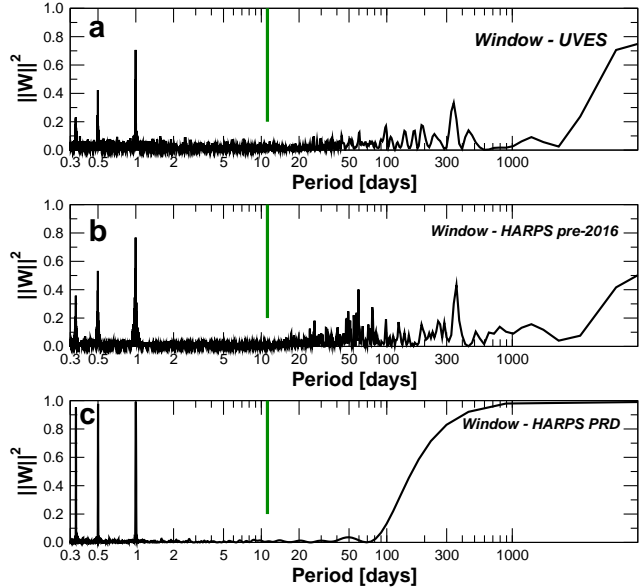

\center
\includegraphics[angle=0, width=0.60\textwidth, clip]{Anglada_EDfig1a.eps}
\includegraphics[angle=0, width=0.60\textwidth, clip]{Anglada_EDfig1b.eps}
\includegraphics[angle=0, width=0.60\textwidth, clip]{Anglada_EDfig1c.eps}
\caption{\textbf{Window function.}  Window function of the UVES (panel a), HARPS
pre-2016(panel b) and  HARPS~PRD (panel c) datasets. The same window function
applies to the time-series of Doppler and activity data. Peaks in the window
function are periods at which aliases of infinite period signals would be
expected.} \label{fig:window} 
\end{figure}

We first present the so-called window function of the three sets under
discussion. The window function is the Fourier transform of the sampling
\cite{dawson:2010}. Its module shows the frequencies (or periods) where a signal
with $0$ frequency (or infinite period) would have its aliases. As shown in
Extended Data Figure \ref{fig:window}, both the UVES and HARPS~PRD campaigns
have a relatively clear window function between 1 and 360 days, meaning that
peaks in periodograms can be interpreted in a very straightforward way (no
aliasing ambiguities). For the UVES case, this happens because the measurements
were uniformly spread over several years without severe clustering, producing
only strong aliases at frequencies beating caused by the usual daily and yearly
sampling (peaks at 360, 1, 0.5 and 0.33 days). The window of the  PRD campaign
is simpler, which is the result of a shorter timespan and the uniform sampling
of the campaign. On the other hand, the HARPS pre-2016 window function (panel b
in Extended Data Figure \ref{fig:window}) contains numerous peaks between 1 and
360 days. This means that signals (e.g.~activity) in the range of a few hundred
days will inject severe interference in the period domain of interest, and
explains why this set is where the Doppler signal at 11.2 days is detected with
less confidence (see Extended Data Figure \ref{fig:per:RV}).

\newpage

\subsection{Radial velocities.}\label{sec:rv}

\begin{figure}[t]
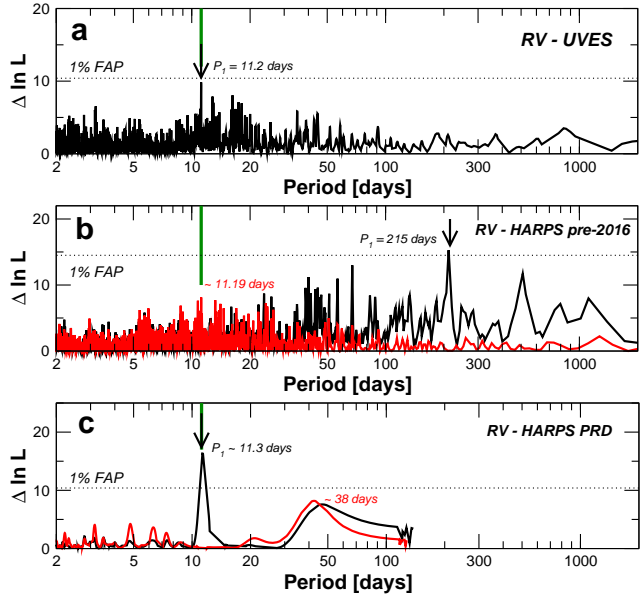

\center
\includegraphics[angle=0, width=0.60\textwidth, clip]{Anglada_EDfig2a.eps}
\includegraphics[angle=0, width=0.60\textwidth, clip]{Anglada_EDfig2b.eps}
\includegraphics[angle=0, width=0.60\textwidth, clip]{Anglada_EDfig2c.eps}
\caption{\textbf{Signal searches on independent radial velocity datasets.}
Likelihood-ratio periodograms searches on the RV measurements of the UVES (panel
a), HARPS pre-2016 (panel b) and HARPS~PRD (panel c) subsets. The periodogram
with all three sets combined is shown in Figure~1 of the main manuscript. Black
and red lines represent the searches for A first and a second signal
respectively.} \label{fig:per:RV}
\end{figure}

Here we present likelihood-ratio periodogram searches for signals in the three
Doppler time-series separately (PRD, HARPS pre-2016, and UVES). They are
analyzed in the same way as the activity indices to enable direct visual
comparison. They differ from the ones presented in the main manuscript in the
sense that they do not include MA terms and the signals are modelled as pure
sinusoids to mirror the analysis of the other time-series as close as possible.
The resulting periodograms are shown in Extended data Figure~\ref{fig:per:RV}. A
signal at 11.2 days was close to detection using UVES data-only. However, let us
note that the signal was not clearly detectable using the Doppler measurements
as provided by the UVES survey\cite{zechmeister:2009}, and it only became
obvious when new Doppler measurements were re-derived using up-to-date Iodine
codes (Section \ref{sec:uves}). The signal is weaker in the HARPS pre-2016
dataset, but it still appears as a possible second signal after modeling the
longer term variability with a Keplerian at 200 days. Sub-sets of the HARPS
pre-2106 data taken in consecutive nights (eg.~HARPS high-cadence runs) also
show strong evidence of the same signal. However splitting the data in subsets
adds substantial complexity to the analysis and the results become quite
sensitive to subjective choices (how to split the data and how to weight each
subset). The combination UVES with all the HARPS pre-2016 (Figure
\ref{fig:periodograms}, panel a) already produced a FAP of $\sim$1\%, but a
dedicated campaign was deemed necessary given the caveats with the sampling and
activity related variability. The HARPS~PRD campaign unambiguously identifies a
signal with the same $\sim 11.2$ days period. As discussed earlier, the
combination of all the data results in a very high significance, which implies
that the period, but also the amplitude and phase are consistent in all three
sets.

\newpage

\subsection{Photometry. Signals and calculation of the FF$^\prime$ index.}
\label{sec:anaphot}

\begin{figure}[t]
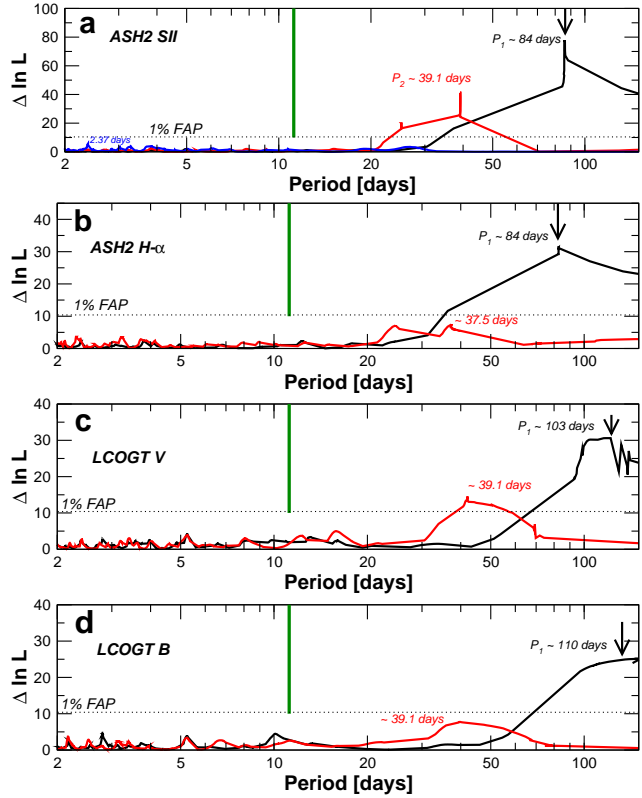

\center
\includegraphics[angle=0, width=0.60\textwidth, clip]{Anglada_EDfig3a.eps}
\includegraphics[angle=0, width=0.60\textwidth, clip]{Anglada_EDfig3b.eps}
\includegraphics[angle=0, width=0.60\textwidth, clip]{Anglada_EDfig3c.eps}
\includegraphics[angle=0, width=0.60\textwidth, clip]{Anglada_EDfig3d.eps}
\caption{\textbf{Signal searches on the photometry}. Likelihood-ratio
periodograms searches for signals in each photometric ASH2 photometric band
(panels a and b) and LCOGT bands (panels c and d). The two sinusoid fit to the
ASH2 SII series ($P_1 = 84$ days, $P_2 = 39.1$ days), is used later to construct
the FF$^\prime$ model to test for correlations of the photometry with the RV
data. Black, red and blue lines represent the search for a first, second and
third signals respectively. }
\label{fig:per:photometry}
\end{figure}

The nightly average of the four photometric series was computed after removing
the measurements clearly contaminated by flares (see Figure~\ref{fig:prddata} in
main manuscript). This produces 43 LCOGT epochs in the B and V bands (80
nights), and 66 ASH2 epochs in both SII and H$_\alpha$ bands (100 nights
covered). The precision of each epoch was estimated using the internal
dispersion within a given night. All four photometric series show evidence of a
long period signal compatible with a photometric cycle at 83-d (likely rotation)
reported before \cite{kiraga2007}. See periodograms in
Extended data Figure~\ref{fig:per:photometry}.

In the presence of spots, it has been proposed that spurious variability should
be linearly correlated with the value of the normalized flux of the star $F$,
the derivative of the flux F$^\prime$, and the product of FF$^\prime$
\cite{aigrain:2012} in what is sometimes called the FF$^\prime$ model. To
include the photometry in the analysis of the Doppler data, we used the best
model fit of the highest quality light curve (AHS2 SII, has the lowest post-fit
scatter) to estimate $F$, $F^\prime$ and $FF^\prime$ at the instant of each PRD
observation. The relation of $F$, $F^\prime$, and $FF^\prime$ to the Doppler
variability is investigated later in the Bayesian analysis of the correlations.

\subsection{Width of the mean spectral line as measured by $m_2$.}\label{sec:m2}
 
\begin{figure}[t]
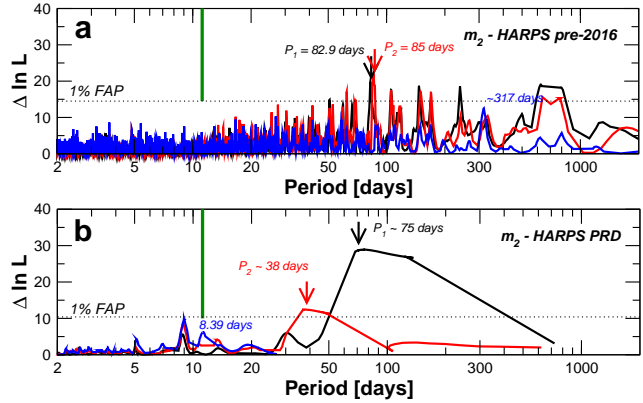

\center
\includegraphics[angle=0, width=0.60\textwidth, clip]{Anglada_EDfig4a.eps}
\includegraphics[angle=0, width=0.60\textwidth, clip]{Anglada_EDfig4b.eps}
\caption{\textbf{Signal searches on the width of the spectral lines.} Likelihood
periodogram searches on the width of the mean spectral line as measured by $m_2$
for the HARPS pre-2016 (panel a) and HARPS~PRD data (panel b). The signals in
the HARPS pre-2016 data are comparable to the photometric period reported in the
literature and the variability in the HARPS~PRD run compares quite well to the
photometric variability. Black, red and blue lines represent the search for a
first, second and third signal respectively.}\label{fig:per:M2} 
\end{figure}

The $m_2$ measurement contains a strong variability that closely mirrors the
measurements from the photometric time-series (see Figure~\ref{fig:prddata} in
the main manuscript). As in the photometry, the rotation period and its first
harmonic ($\sim 40$ days) are clearly detected in the PRD campaign (see Extended
data Figure~\ref{fig:per:M2}). This apparently good match needs to be verified
on other stars as it might become a strong diagnostic for stellar activity in
M-stars. The analysis of the HARPS pre-2016 also shows very strong evidence that
$m_2$ is tracing the photometric rotation period of 83 days. The modelling of
this HARPS pre-2016 requires a second sinusoid with $P_2 \sim 85$~days, which is
peculiar given how close it is to $P_1$. We suspect this is caused by
photospheric features on the surface changing over time.

\subsection{Asymmetry of the mean spectral lines as monitored by $m_3$}
\label{sec:m3}

\begin{figure}[t]
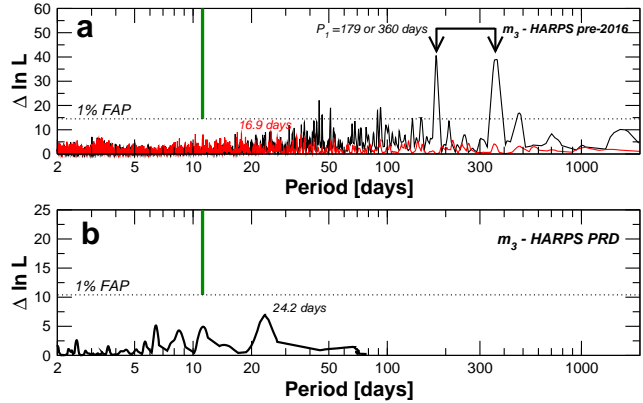

\center
\includegraphics[angle=0, width=0.60\textwidth, clip]{Anglada_EDfig5a.eps}
\includegraphics[angle=0, width=0.60\textwidth, clip]{Anglada_EDfig5b.eps}
\caption{\textbf{Signal searches on the assymetry of the spectral lines.}
Likelihood periodogram searches on the line asymmetry as measured by
$m_3$ from the HARPS pre-2016 (panel a) and HARPS~PRD (panel b) datasets. A signal
beating at $\sim$ 1 year and 1/2 year is detected in the HARPS pre-2016 data,
possibly related to instrumental systematic effects or telluric contamination. 
No signals are detected above 1\% threshold in the HARPS~PRD campaign. Black and 
red lines represent the search for first and second signals respectively. }
\label{fig:per:M3}
\end{figure}

The periodogram analysis of $m_3$ of the PRD run suggests a signal at 24 days
which is close to twice the Doppler signal of the planet candidate (see Extended
Data Figure~\ref{fig:per:M3}). However, line asymmetries are expected to be directly
correlated with Doppler signals, not at twice nor integer multiples of the
Doppler period. In addition, the peak has a FAP$\sim$ 5\% which makes it
non-significantly different from white noise. When looking at the HARPS pre-2016
data, some strong beating is observed at 179 and 360 days, which is likely
caused by a poorly sampled signal at that period or longer (magnetic cycle?), or
some residual systematic effect (contamination by tellurics?). In summary, $m_3$
does not show evidence of any stable signal in the range of interest.

\subsection{Signal searches in S-index.} \label{sec:searchsindex}

\begin{figure}[t]
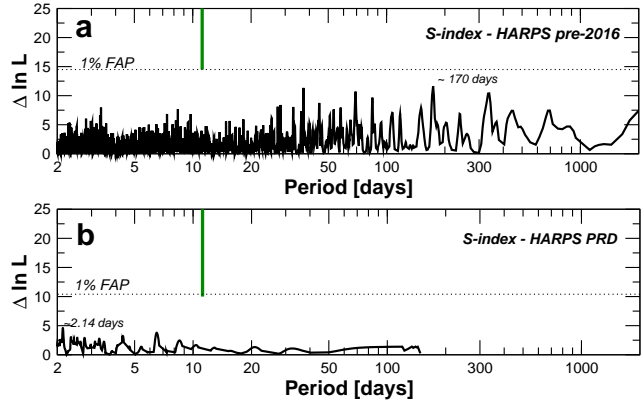

\center
\includegraphics[angle=0, width=0.60\textwidth, clip]{Anglada_EDfig6a.eps}
\includegraphics[angle=0, width=0.60\textwidth, clip]{Anglada_EDfig6b.eps}
\caption{\textbf{Signal searches on the chromospheric S-index.}
Likelihood-ratio periodogram of S-index from the HARPS~pre-2016 (panel a)
and HARPS~PRD (panel b) campaigns. No signals detected above 1\% threshold.}
\label{fig:per:SINDEX}
\end{figure}

While H$_\alpha$ \cite{robertson:2014} and other lines like the sodium doublet
(NaD1 and NaD2) \cite{silva:2012} have been shown to be the best tracers for
activity on M-dwarfs, analyzing the time-series of the S-index is also useful
because of its historical use in long term monitoring of main-sequence stars
\cite{baliunas:1995}. In Extended Data Figure~\ref{fig:per:SINDEX} we show the
likelihood ratio periodograms for the $S$-indices of the HARPS pre-2016 and PRD
time-series. As can be seen, no signals were found around the 11~day period of
the radial velocity signal, however two peaks were found close the 1\% false
alarm probability threshold with periods of $\sim$170 and 340~days. In order to
further test the reality of these possible signals, we performed a Lomb-Scargle
(LS) periodogram analysis \cite{scargle:1982} of the combined PRD and pre-2016
HARPS data.  This test resulted in the marginal recovery of both the 170 and 340
day peaks seen in the likelihood periodograms, with no emerging peaks around the
proposed 11~day Doppler signal.  The LS tests revealed some weak evidence for a
signal at much lower periods, $\sim$7~days and $\sim$30~days.

Given that there is evidence for significant peaks close to periods of 1~yr, its
first harmonic, and the lunar period, we also analysed the window function of
the time-series to check if there was evidence that these peaks are artefacts
from the combination of the window function pattern interfering with a real
long-period activity signal in the data. The dominant power in the window
function is found to increase at periods greater than 100~days, with a forest of
strong peaks found in that domain, in comparison to sub-100 day periods which is
very flat, representing the noise floor of the time-series. This indicates that
there is likely to be strong interference patterns from the sampling in this
region, and that the signal in the radial velocity data is also not due to the
sampling of the data. A similar study in the context of the HARPS M-dwarf
program was also done on Proxima \cite{silva:2012}. They compared several
indices and finally decided to use the intensity of the chromospheric sodium
doublet lines. They did not report any significant period at the time, but we
suspect this was due to using fewer measurements, and not removing the
frequent flaring events from the series, which also requires compilation of a
number of observations to reliably identify outliers caused by flares.
 
\subsection{Signal searches in H$_\alpha$ emission}\label{sec:searchhalpha}

\begin{figure}[t]
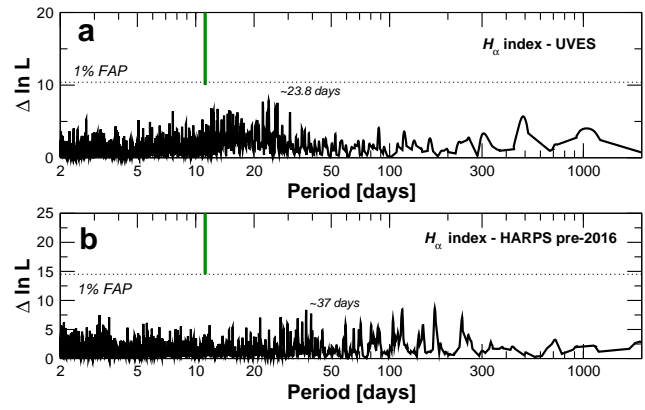

\center
\includegraphics[angle=0, width=0.60\textwidth, clip]{Anglada_EDfig7a.eps}
\includegraphics[angle=0, width=0.60\textwidth, clip]{Anglada_EDfig7b.eps}
\caption{\textbf{Signal searches on the spectroscopic H$_\alpha$ index}
Likelihood-ratio periodogram searches of $H_\alpha$ intensity from
the UVES (panel a), HARPS pre-2016 (panel b) and HARPS~PRD (panel c) campaigns. No signals
detected above 1\% threshold.}
\label{fig:per:HALPHA}
\end{figure}

Our likelihood-ratio periodograms for $H_\alpha$ (Extended Data
Figure~\ref{fig:per:HALPHA}) only show low significance peaks in the 30-40 days
period range. It is important to note that the analyses described above have
been performed on multiple versions of the dataset, in the sense that we
analysed the full dataset without removing measurements affected by flaring,
then proceeded to reanalyse the activities by dropping data clearly following
the flaring periods that Proxima went through when we observed the star. This
allowed us to better understand the impact that flares and outliers have on
signal interference in the activity indices. Although the distribution of peaks
in periodograms changes somewhat depending on how stringent the cuts are, no
emerging peaks were seen close to an 11~day period. Concerning UVES H$_\alpha$
measurements, our likelihood-ratio periodogram did not detect any significant
signal.

\subsection{Further tests on the signal.}\label{sec:extrasignal}
 
It has been shown \cite{pascual:2015} that at least some of the ultraprecise
photometric time-series measured by CoRot and Kepler space missions do not have
a necessary property to be represented by a Fourier expansion: the underlying
function, from which the observations are a sample, must be analytic. An
algorithm introduced in the same paper can test this property and was applied to
the PRD data. The result is that, contrary to the light curves aforementioned,
claims that the underlying function is non-analytic does not hold with the
information available. Though the null hypothesis cannot be definitively
rejected, at least until more data is gathered, our results are consistent with
the hypothesis that a harmonic component is present in the Doppler time-series.
 
\newpage 
 
\subsection{Flares and radial velocities.}\label{sec:search:rv}

\begin{figure}[t]
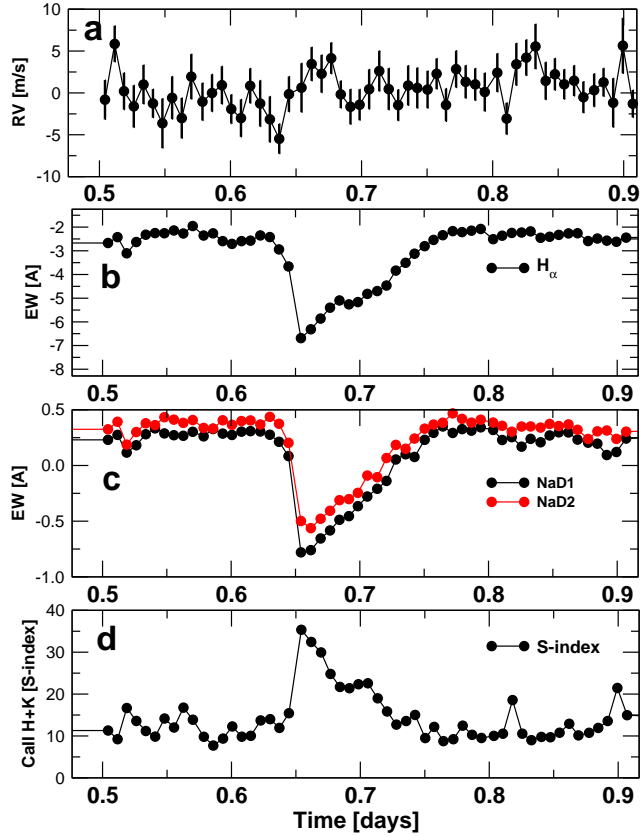

\center
\includegraphics[angle=0, width=0.60\textwidth, clip]{Anglada_EDfig8a.eps}
\includegraphics[angle=0, width=0.60\textwidth, clip]{Anglada_EDfig8b.eps}
\caption{\textbf{Radial velocities and chromospheric emission
during a flare.}
Radial velocities (panel a) and equivalent width measurements of the
H$_\alpha$ (panel b), Na Doublet lines (panel c), and the S-index (panel d) 
as a function of time during a flare that occurred the night of May 5th, 2013. 
Time axis is days since JD=245417.0 days. No trace of the flare is observed 
on the RVs.}
\label{fig:flare} 
\end{figure}

Among the high-cadence data from May 2013 with HARPS, two strong flares are
fully recorded. During these events, all chromospheric lines become prominent
in  emission, H$_\alpha$ being the one that best traces the characteristic
time-dependence of flares observed on other stars and the Sun. The spectrum and
impact of flares on the RVs will be described elsewhere in detail. Relevant to
this study, we show th    at the typical flares on Proxima do not produce
correlated Doppler shifts (Extended Data Figure~\ref{fig:flare}). This justifies
the removal of obvious flaring events when investigating signals and
correlations in the activity indices.

\newpage 
\section{Complete model and Bayesian analysis of the activity coefficients.}
\label{sec:bayesanalysis}
 
\begin{figure}[t] 
\center
\includegraphics[angle=0, width=0.95\textwidth]{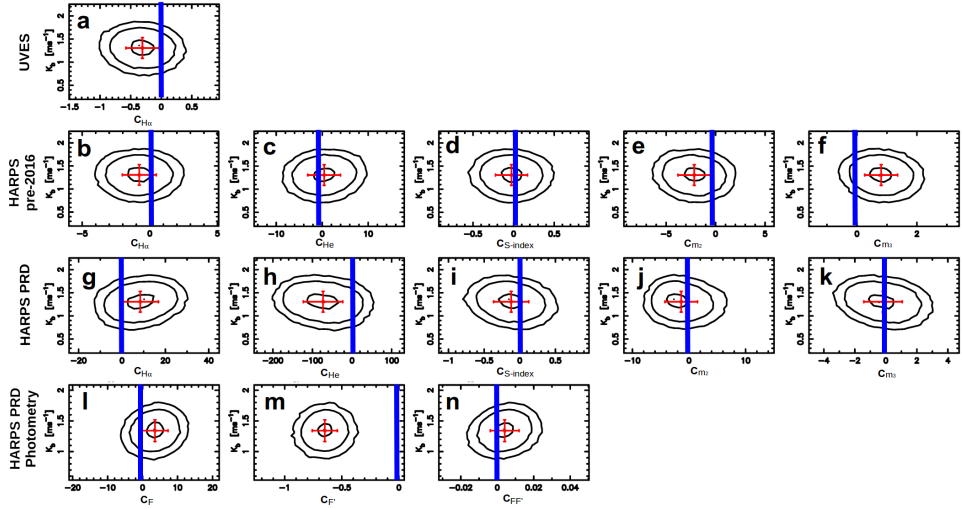}
\caption{\textbf{Probability distributions for the activity
coefficients  versus signal amplitude.} Marginalized posterior densities of the
activity coefficients  versus the semi-amplitude of the signal for 
UVES (panel a), HARPS pre-2016 (panels b,c,d,e,f), HARPS~PRD campaign (panels g,h,i,j,k) 
and the
photometric FF$^\prime$ indices for the PRD campaign only (panels 
l, m, n). Each panel shows equiprobability contours
containing 50\%, 95\%, and 99\% of the probability density around the mean
estimate, and the corresponding standard deviation of the marginalized
distribution (1-$\sigma$) in red. The blue bar shows the zero value of each
activity coefficient. Only C$_{F^\prime}$ is found to be significantly
different from zero.}
\label{fig:cor}
\end{figure}

A global analysis including all the RVs and indices was performed to verify that
the inclusion of correlations would reduce the model probability below the
detection thresholds. Equivalently, the Doppler semi-amplitude would become
consistent with zero if the Doppler signal was to be described by a linear
correlation term. Panels in Extendent Data Figure~\ref{fig:cor} show marginalized
distributions of linear correlation coefficients with the Doppler semi-amplitude
$K$. Each subset is treated as a separate instrument and has its own zero-point,
jitter and Moving Average term (coefficient) and its activity coefficients. In
the  final model, the time-scales of the Moving Average terms are fixed to
$\sim$ 10 days because they were not contrained within the prior bounds, thus
compromising the convergence of the chains. The sets under consideration are

\begin{itemize}

\item \textbf{UVES} : 70 radial velocity measurements and corresponding
H$_\alpha$ emission measurements.

\item \textbf{HARPS pre-2016} : 90 radial velocity measurements obtained between
2002 and 2014 by several programmes and corresponding spectroscopic indices :
$m_2$, $m_3$, S-index, and the intensities of the H$_\alpha$ and HeI lines as
measured on each spectrum.

\item \textbf{HARPS PRD} : 54 Doppler measurements obtained between Jan 18th-Mar
31st, 2016, and the same spectroscopic indices as for the HARPS pre-2016. The
values of the F,  F$^\prime$ and FF$^{\prime}$ indices were obtained by
evaluating the best fit model to the ASH2 SII  photometric series at the HARPS
epochs  (see Section \ref{sec:anaphot}).

\end{itemize}

An activity index is correlated with the RV measurements in a given set if the
zero value of its activity coefficient is excluded from the 99\% credibility
interval. Extended Data Figure~\ref{fig:cor} shows the equiprobability contours
containing 50\%, 95\%, and 99\% of the probability density around the mean
estimate, and the corresponding 1-$\sigma$ uncertainties in red. Only the
$F^\prime$ index (time derivative of the photometric variability) is
significantly different from 0 at high confidence (Extended Data
Figure~\ref{fig:cor}, bottom row, panel m). Linking this correlation to a
physical process requires further investigation. To ensure that such
correlations are causally related, one needs a model of the process causing the
signal in both the RV and the index, and in the case of the photometry one would
need to simultaneously cover more stellar photometric periods to verify that the
relation holds over time. Extended Data Table~\ref{tab:allpars} contains a
summary of all the free parameters in the model including activity coefficients
for each dataset.

%\input{main.bbl2}
%\bibliography{extra/biblio}

\begin{table}[t]
\begin{minipage}{\textwidth}
 
\caption{\textbf{Complete set of model parameters}. The
definition of all the parameters is given in Section~\ref{sec:statistics} of the
methods. The values are the maximum \emph{a posteriori} estimates and the
uncertainties are expressed as 68\% credibility intervals.
The reference epoch for this solution is Julian Date 
$t_0=2451634.73146$ days, which corresponds to the first UVES epoch.
$^*$Units of the activity coefficients are \ms divided by the units of
each activity index.
}\label{tab:allpars}

\begin{center}
\small
\begin{tabular}{lll}
\hline \hline
Parameter &  Mean [68\% c.i.] & Units                    \\
\hline
Period           & 11.186 [11.184, 11.187] & days \\
Doppler Amplitude  & 1.38 [1.17, 1.59] & ms$^{-1}$             \\
Eccentricity  & $<$0.35  & --                      \\
Mean Longitude & 110 [102, 118] & deg   \\
Argument of periastron   & 310 [-]  & deg                       \\
Secular acceleration   & 0.086 [-0.223, 0.395] & ms$^{-1}$yr$^{-1}$\\
\\
Noise parameters\\
\hline
$\sigma_{\rm HARPS}$  & 1.76 [1.22, 2.36] & ms$^{-1}$ \\
$\sigma_{\rm PRD}$  & 1.14 [0.57, 1.84] & ms$^{-1}$\\
$\sigma_{\rm UVES}$  & 1.69 [1.22, 2.33] & ms$^{-1}$ \\
$\phi_{\rm HARPS}$  & 0.93 [0.46, 1] & ms$^{-1}$\\
$\phi_{\rm PRD}$    & 0.51 [-0.63, 1] & ms$^{-1}$ \\
$\phi_{\rm UVES}$   & 0.87 [-0.02, 1] & ms$^{-1}$\\
\\
Activity coefficients\footnote{Units of the activity coefficients are \ms divided by the units of
each activity index.}\\
\hline
\\
UVES\\
$C_{\rm H\alpha}$ & -0.24 [-1.02, 0.54] \\
\\
HARPS pre-2016\\
$C_{\rm H\alpha}$ & -0.63 [-4.13, 3.25] \\
$C_{\rm He}$ & 1.0 [-9.3, 11.4] \\
$C_{\rm S}$ & -0.027 [-0.551, 0.558] \\
$C_{\rm m_2}$ & -1.93 [-6.74, 2.87] \\
$C_{\rm m_3}$ & 0.82 [-0.60, 2.58] \\
\\
HARPS PRD\\
$C_{\rm H\alpha}$ & 9.6 [-12.9, 33.3] \\
$C_{\rm He}$ & -77 [-210, 69] \\
$C_{\rm S}$ & -0.117 [-0.785, 0.620] \\
$C_{\rm m_2}$ & -2.21 [-8.86, 7.96] \\
$C_{\rm m_3}$ & -0.02 [-3.67, 3.44] \\
\\
PRD photometry\\
$C_{\rm F}$ & 0.0050 [-0.0183, 0.0284] \\
$C_{\rm F'}$ & -0.633 [-0.962, -0.304] \\
$C_{\rm FF'}$ & 4.3 [-6.8, 14.8] \\
\\
\end{tabular}
\end{center}
 
\end{minipage}
\end{table}

\end{document}